\documentclass[conference]{IEEEtran}
\IEEEoverridecommandlockouts
\usepackage{cite}
\usepackage{amsmath,amssymb,amsfonts}
\usepackage{graphicx}
\usepackage{textcomp}
\usepackage{xcolor}

\usepackage{algorithm}
\usepackage{algcompatible}

\algnewcommand\algorithmicreturn{\textbf{return}}
\algnewcommand\RETURN{\State \algorithmicreturn}%
\usepackage[noend]{algpseudocode}
\mathchardef\mhyphen="2D
\algdef{SE}[DOWHILE]{Do}{doWhile}{\algorithmicdo}[1]{\algorithmicwhile\ #1}%
\usepackage{mathtools}
\DeclarePairedDelimiter\ceil{\lceil}{\rceil}

\def\BibTeX{{\rm B\kern-.05em{\sc i\kern-.025em b}\kern-.08em
    T\kern-.1667em\lower.7ex\hbox{E}\kern-.125emX}}
\begin{document}

\title{Prioritized Variable-length Test Cases Generation for Finite State Machines}

\author{\IEEEauthorblockN{Vaclav Rechtberger}
\IEEEauthorblockA{\textit{Dept. of Computer Science, FEE} \\
\textit{Czech Technical University in Prague}\\
Prague, Czechia \\
rechtva1@fel.cvut.cz}
\and
\IEEEauthorblockN{Miroslav Bures}
\IEEEauthorblockA{\textit{Dept. of Computer Science, FEE} \\
\textit{Czech Technical University in Prague}\\
Prague, Czechia \\
miroslav.bures@fel.cvut.cz }
\and
\IEEEauthorblockN{Bestoun S. Ahmed}
\IEEEauthorblockA{\textit{Dept. of Mathematics and Computer Science} \\
\textit{Karlstad University, Sweden \&} \\ Czech Technical University in Prague \\
bestoun@kau.se}
\and
\IEEEauthorblockN{Youcef Belkhier}
\IEEEauthorblockA{\textit{Dept. of Computer Science, FEE} \\
\textit{Czech Technical University in Prague}\\
Prague, Czechia \\
youcef.belkhier@fel.cvut.cz}
\and
\IEEEauthorblockN{Jiri Nema}
\IEEEauthorblockA{\textit{Faculty of Military Health Sciences} \\
\textit{University of Defence}\\
Hradec Kralove, Czechia \\
jiri.nema@unob.cz}
\and
\IEEEauthorblockN{Hynek Schvach}
\IEEEauthorblockA{\textit{Faculty of Military Health Sciences} \\
\textit{University of Defence}\\
Hradec Kralove, Czechia \\
hynek.schvach@unob.cz}
}

\maketitle

\begin{abstract}
Model-based Testing (MBT) is an effective approach for testing when parts of a system-under-test have the characteristics of a finite state machine (FSM). Despite various strategies in the literature on this topic, little work exists to handle special testing situations. More specifically, when concurrently: (1) the test paths can start and end only in defined states of the FSM, (2) a prioritization mechanism that requires only defined states and transitions of the FSM to be visited by test cases is required, and (3) the test paths must be in a given length range, not necessarily of explicit uniform length. This paper presents a test generation strategy that satisfies all these requirements. A concurrent combination of these requirements is highly practical for real industrial testing. Six variants of possible algorithms to implement this strategy are described. Using a mixture of 180 problem instances from real automotive and defense projects and artificially generated FSMs, all variants are compared with a baseline strategy based on an established N-switch coverage concept modification. Various properties of the generated test paths and their potential to activate fictional defects defined in FSMs are evaluated. The presented strategy outperforms the baseline in most problem configurations. Out of the six analyzed variants, three give the best results even though a universal best performer is hard to identify. Depending on the application of the FSM, the strategy and evaluation presented in this paper are applicable both in testing functional and non-functional software requirements. 
\end{abstract}



\section{Introduction}

\color{blue}\textbf{Paper accepted at the 6th International Workshop on Testing Extra-Functional Properties and Quality Characteristics of Software Systems (ITEQS) workshop of the 15th IEEE International Conference on Software Testing, Verification and Validation (ICST) 2022 conference, April 4, 2022 – April 13, 2022, https://icst2022.vrain.upv.es/ }
\color{black}

\hspace{3mm}

Finite State Machine (FSM) character is available in various parts of most of the modern systems, like software, electronics, or the Internet of Things (IoT). To test these systems, a usual approach is to exercise sequences of transitions (test paths) in FSMs to model those parts of the System Under Test (SUT) \cite{ammann2016introduction}. The goal is to create test paths (test path set) in an optimal way, where various criteria can be considered, e.g., minimal test price or maximal probability to detect defects in SUT. The satisfaction of a defined test coverage criteria is a standard prerequisite of such a process \cite{utting2012taxonomy,ahmad2019model}. The Model-based Testing (MBT) approach renders as an effective automated way to achieve this goal \cite{ammann2016introduction,utting2012taxonomy,ahmad2019model}.


Current literature presents a variety of approaches to generate test paths from an FSM-based SUT model (e.g., \cite{lee1996principles,aggarwal2012test,ammann2016introduction,dias2007survey}). However, fewer works exist for specific cases, in which the following requirements have to be satisfied concurrently: (1) test paths can start and end in defined FSM states, which are a subset of all FSM states, (2) test paths must visit only defined priority states and transitions of FSM; the rest of the states and transitions are not required to be visited by a test path set, and, (3) the length of the created test paths must be in a defined range (not explicitly of a certain uniform length). Concurrent satisfaction of these three requirements gives an FSM-based testing technique good flexibility and potential to serve the needs of industry projects. Formulation of the requirements results from our active discussions with industry testing experts in Rockwell, Siemens, Skoda Auto and Electrolux companies, in which the needs for an FSM-based testing technique were analyzed.

In practical testing, starting and ending a test in certain states of FSM either is costly or not possible \cite{derderian2010estimating,kalaji2009generating}. The length of the test paths also plays a significant role. Excessively long test paths might suffer from high maintenance. Additionally, if a test path is interrupted by a defect present in a SUT, it might be challenging to finish the test properly. Opposite extreme, too short test paths might be burdened by overhead that is caused by test initialization clean-up after the test of related managerial efforts \cite{vroon2013tmap}.


Hence, this paper proposes the Prioritized State Machine Test (PSMT) strategy that generates sets of test paths that satisfy the aforementioned requirements. The strategy employs an FSM-based SUT model and two types of test coverage criteria to model this problem. As a part of the strategy, we present six algorithm variants that generate the test paths to satisfy the given test coverage criteria. As a baseline, we use an alternative N-Switch Reduction algorithm that is based on an established N-switch coverage concept \cite{vroon2013tmap,pol2002software,aalst2008tmap}. We compare the results of the presented algorithms using 180 SUT models that are partially derived from real automotive and defense systems, and the parts are generated artificially.


\section{SUT model}
\label{sec:proposed_sut_model}

The SUT model $\boldsymbol{G} = (V,E,L,\varepsilon,v_s,V_e,V_{ts},V_{te})$ is a directed multigraph, where $V$ is a set of vertices representing FSM states, $E$ is a set of edges representing FSM transitions, and $L$ is a set of edge labels, edge $e \in E$ defined by $\varepsilon:~E\rightarrow{}~\{(s, f, l)~|~s, f \in V \land l \in L\}$, where $s$ is a start vertex of edge $e$, $f$ is an end vertex of edge $e$, and $l$ is a label of edge $e$.

Vertex $v_s \in V$ is a start of the state machine, $V_e \subset V$ is a set of end vertices of the state machine, $V_{ts} \subset V$ is a set of possible starts of test paths, $V_{te} \subset V$ is a set of possible end of test paths, $v_s \in V_{ts}$ and $V_{e} \subset V_{te}$ and $V_{ts} \cap V_{te}$ can be nonempty. Each edge $e \in E$ has defined its priority denoted as $priority(e)$ as well as each vertex $v \in V$ has defined its priority denoted as $priority(v)$. This priority is a real number and it holds that $minPriority \leq priority(e) \leq maxPriority$ and $minPriority \leq priority(v) \leq maxPriority$, where $minPriority$ and $maxPriority$ denote minimal and maximal level of priority, $minPriority < maxPriority$.

Test path $p$ is a path in $\boldsymbol{G}$ that starts in a $v_{ts} \in V_{ts}$ and ends in a $v_{te} \in V_{te}$. A test path is a sequence of edges and $P$ is a set of all test paths. 


\section{Test coverage criteria}
\label{sec:test_coverage_criteria}

We define two test coverage criteria that must be satisfied in $P$. They differ by the number of test path transitions, giving additional flexibility to generate a test path set suitable for a particular case.

$P$ satisfies  \textbf{PSMT Basic Coverage}, when all the following conditions are satisfied:
\begin{enumerate}
\item Each of test paths $p \in P$ must start in a vertex from $\{v_s\} \cup V_{ts}$ and end in a vertex from $V_e \cup V_{te}$,

\item for each $p \in P$, $minLenght \leq length(p) \leq maxLength$, where $length(p)$ is length of a test path $p$ in number of its edges,

\item each vertex $v \in V$ and edge $e \in E$ with $priority(v) \geq priority\_threshold$ and $priority(e) \geq priority\_threshold$ that can be part of some $p \in P$ which starts at some vertex from $\{v_s\} \cup V_{ts}$, ends in some vertex from $V_e \cup V_{te}$ and $minLenght \leq length(p) \leq maxLength$, where $length(p)$ is length of $p$ in number of its edges, must be present in $p$, and,

\item any path $p \in P$ is not a sub-path of other paths in $P$.

\end{enumerate}

On top of these rules, no additional requirements on how $V_{ts}$ and $V_{te}$ have to be chained or combined in the test paths are given by \textit{PSMT Basic Coverage}. Further, it is not required that every vertex from $V_e \cup V_{te}$ must be present as an end vertex of some $p \in P$. To satisfy \textit{PSMT Basic Coverage}, it is not necessary to visit neither whole $E \in \boldsymbol{G}$ nor even $V \in \boldsymbol{G}$.

The \textit{PSMT Basic Coverage} is suitable for lower intensity tests when the situation does not require to conduct intensive testing or there is a lack of testing resources.

$P$ satisfies \textbf{PSMT Extended Coverage}, when all the following conditions are satisfied:
\begin{enumerate}
\item $P$ satisfies \textit{PSMT Basic Coverage}, and,


\item each edge incoming and outgoing to each vertex $v \in V$ with $priority(v) \geq priority\_threshold$ must be present in at least one path $p \in P$, and,
\item each pair of adjacent edges $(e_1,e_2), e_1 \in E, e_2 \in E$, in which $priority(e_1) \geq priority\_threshold$ or $priority(e_2) \geq priority\_threshold$, must be part of some $p \in P$.



\end{enumerate}

When $P$ satisfies \textit{PSMT Extended Coverage}, it satisfies also \textit{PSMT Basic Coverage} criterion. The \textit{PSMT Extended Coverage} is designed for more intensive tests and more transitions of FSM are executed during the tests. However, to satisfy \textit{PSMT Extended Coverage}, it is not required to visit all transitions from $E \in \boldsymbol{G}$ nor all states $V \in \boldsymbol{G}$.

A consequence of \textit{PSMT Basic Coverage} and \textit{PSMT Extended Coverage} is that for a certain $\boldsymbol{G}$ combined with certain ranges of $minLenght$ and $maxLenght$, $P$ that satisfies the test coverage criterion might not exist. This situation can be solved by changing $minLenght$ and $maxLenght$, adding more $V_{ts}$ and $V_{te}$ to $\boldsymbol{G}$, or by a combination of both.


\section{Algorithms}
\label{sec:algorithms}

PSMT strategy is defined by Algorithm \ref{alg:GenerateTestPaths}. It accepts SUT model $\boldsymbol{G}$, $minPriority$, $maxPriority$, expected test path length range $minLength$ and $maxLength$, test coverage criterion denotes as $testCoverage$ that can be \textit{PSMT Basic Coverage} and \textit{PSMT Extended Coverage}, and test path set reduction type denoted as $reductionType$ that determines the strategy variant. Requirement $r \in R$ is a path in $\boldsymbol{G}$ or a vertex $v \in V$ (a zero length path) that must be present in a $p \in P$. Moreover, each $p \in P$ contains at least one $r \in R$. Moreover, $R$ must satisfy criteria as defined in Section \ref{sec:test_coverage_criteria} for selected $testCoverage$.

Algorithm \ref{alg:FindShortestPathInRange} is a subroutine of the PSMT strategy and creates an initial set of test paths $P$ that are further reduced by the following algorithms. As the test path set reduction type (specified by $reductionType$), we analyzed six options, starting with \textit{No reduction} variant, followed by five reduction variants as defined by separate algorithms \ref{alg:RandomTestPathsReduction}, \ref{alg:SortedTestPathsReduction}, \ref{alg:ChvatalsImprovedTestPathsReduction}, \ref{alg:GeneticTestPathReduction} and \ref{alg:SimulatedAnnealingTestPathReduction}. 

These algorithms approach the problem of test path set reduction as a well-known Set Cover Problem (SCP), where a set of requirements $R$ is an universe and $P$ is as a set of subsets of this universe, where each path $p \in P$ represents a subset of requirements $R_c \subseteq R$ it covers.

\textbf{No reduction} is a trivial variant that does not reduce $P$.

\begin{algorithm}
\caption{Generate $P$ for $\boldsymbol{G}$ by PSMT strategy}\label{alg:GenerateTestPaths}
\hspace*{\algorithmicindent} \textbf{\bf Function: GenerateTestPathsPSMT}\\
\hspace*{\algorithmicindent} \textbf{Input:} $\boldsymbol{G}$, $minLength$, $maxLength$,
$testCoverage$, $minPriority$, $maxPriority$, $reductionType$ \\
\hspace*{\algorithmicindent} \textbf{Output:} Set of test paths $P$
\begin{algorithmic}[1]
\State $P \gets \emptyset$  \Comment{an empty set of test paths}

\State $R \gets$ generate requirements for $testCoverage$ for vertices $v \in V$ with $minPriority \leq priority(v) \leq maxPriority$ and edges $e \in E$ with $minPriority \leq priority(e) \leq maxPriority$ by specification in Section \ref{sec:test_coverage_criteria}.

\For{$r \in R$}
    \State $p \gets$ \textbf{FindShortestPathInRange(}$r$\textbf{)} \Comment{Requirement $r$ is a path in $\boldsymbol{G}$ or vertex $v \in V$}
    \State $P \gets P \cup \{p\}$
\EndFor

\State $P \gets P_{reduced}$ returned by Test Path Reduction algorithm accepting $P$ and $R$ that is specified by $reductionType$ (no reduction or Algorithms \ref{alg:RandomTestPathsReduction}, \ref{alg:SortedTestPathsReduction}, \ref{alg:ChvatalsImprovedTestPathsReduction}, \ref{alg:GeneticTestPathReduction}, \ref{alg:SimulatedAnnealingTestPathReduction}). 
\State \textbf{return} $P$
\Statex{\bf end function}
\end{algorithmic}
\end{algorithm}

\begin{algorithm}
\caption{Find the shortest path in range}\label{alg:FindShortestPathInRange}
\hspace*{\algorithmicindent} \textbf{Function: FindShortestPathInRange}\\
\hspace*{\algorithmicindent} \textbf{Input:} $requirement$, $minLength$, $maxLength$\\
\hspace*{\algorithmicindent} \textbf{Output:} Path $p$
\Comment{if no path is found then an empty path is returned}
\begin{algorithmic}[1]
    \State $E_{map}$ and $S_{map}$ are empty maps of paths. Key in the map is a path length. 
    \State $E_{queue}$ and $S_{queue}$ are empty queues of paths
    \State $p_{start}, p_{end} \gets$ first and last vertex of $requirement$ 
    \State \textbf{push } $p_{end}$ to $E_{queue}$; 
    \State \textbf{push } $p_{start}$ to $S_{queue}$
    
    \While{($E_{queue}$ is not empty) $\lor$ ($S_{queue}$ is not empty)}
        \If{$S_{queue}$ is not empty}
            \State $p_{start} \gets$ \textbf{pull from } $S_{queue}$
            \If{$|p_{start}| <= maxLength$}
                \State $v_{first} \gets$ first vertex of $p_{start}$
                \If{$v_{first} \in V_{ts}$}
                    \If{$E_{map}$ contains a path that can be joined with $p_{start}$ so that for the result path $x$, $minLength \leq length(x) \leq minLength$}
                        \State \textbf{return} $p_{start}$ concatenated by the path from $E_{map}$
                    \EndIf
                    \If{$|p_{start}| +$ length of shortest path from $E_{map} \leq maxLength$}
                        \If{ $S_{map}[|p_{start}|]$ does not exist}
                            \State $S_{map}[|p_{start}|] \gets p_{start}$
                        \EndIf
                    \EndIf
                \EndIf
                \If{$|p_{start}| + 1~+ $ length of the shortest path from $E_{map} \leq maxLength$}
                    \For{$e \in $ edges incoming to $v_{first}$}
                        \State \textbf{put} ($e$ \textbf{concatenate} $p_{start}$) to $S_{queue}$
                    \EndFor
                \EndIf
            \EndIf
        \EndIf
        \If{$E_{queue}$ is not empty}
            \State $p_{end} \gets$ \textbf{pull from } $E_{queue}$
            \If{$|p_{end}| <= maxLength$}
                \State $v_{last} \gets$ last vertex of $p_{end}$
                \If{$v_{last} \in V_{te}$}
                    \If{$S_{map}$ contains a path that can be joined with $p_{end}$ so that for the result path $x$, $minLength \leq length(x) \leq minLength$}
                        \State \textbf{return} the path from $S_{map}$ concatenated by $p_{end}$
                    \EndIf
                    \If{$|p_{end}| +$ length of shortest path from $S_{map} \leq maxLength$}
                        \If{$E_{map}[|p_{end}|]$ does not exist}
                            \State $E_{map}[|p_{end}|] \gets p_{end}$
                        \EndIf
                    \EndIf
                \EndIf
                \If{$|p_{end}| + 1~+ $ length of the shortest path from $S_{map} \leq maxLength$}
                    \For{$e \in $ edges outgoing from $v_{last}$}
                        \State \textbf{put} ($p_{start}$ \textbf{concatenate} $e$) to $E_{queue}$
                    \EndFor
                \EndIf
            \EndIf
        \EndIf
    \EndWhile
\State \textbf{return} empty path
\Statex{\bf end function}
\end{algorithmic}
\end{algorithm}

The simplest approach is \textbf{Random} specified by Algorithm \ref{alg:RandomTestPathsReduction}. It iterates over all $p \in P$ until all $r \in R$ are covered. If the iterated $p$ covers some uncovered requirements, they are marked as covered and the path is added to the final set $P_{reduced}$, otherwise iterating continues. Finally, the reduced set $P$ is returned.

\begin{algorithm}
\caption{Random $P$ reduction}\label{alg:RandomTestPathsReduction}
\hspace*{\algorithmicindent} \textbf{\bf Function: Random}\\
\hspace*{\algorithmicindent} \textbf{Input:} $P$, $R$\\
\hspace*{\algorithmicindent} \textbf{Output:} Reduced set of test paths $P_{reduced}$
\begin{algorithmic}[1]
\State $R_{uncovered} \gets R$
\State $P_{iterator} \gets P$
\State $P_{reduced} \gets$ an empty set of paths

\While {$R_{uncovered} \neq \emptyset \land P_{iterator} \neq \emptyset$}
    \State $p_{next} \gets$ a random path from $P_{iterator}$
    \State $P_{iterator} \gets P_{iterator} \setminus \{p_{next}\}$
    \State $R_{covered} \gets$ previously uncovered requirements now covered by $p_{next}$
    \If{$R_{covered} \neq \emptyset$}
        \State $R_{uncovered} \gets R_{uncovered} \setminus R_{covered}$
        \State $P_{reduced} \gets P_{reduced} \cup \{p_{next}\}$
    \EndIf
\EndWhile

\State \textbf{return} $P_{reduced}$
\Statex{\bf end function}
\end{algorithmic}
\end{algorithm}

The variant \textbf{Sorted} is defined in Algorithm \ref{alg:SortedTestPathsReduction}. The test path set $P$ is first sorted in a descendant order by a number of covered requirements. The algorithm continues as in the random variant (see Algorithm \ref{alg:RandomTestPathsReduction}), but paths are taken in sorted order.

\begin{algorithm}
\caption{Sorted $P$ Reduction}\label{alg:SortedTestPathsReduction}
\hspace*{\algorithmicindent} \textbf{\bf Function: Sorted}\\
\hspace*{\algorithmicindent} \textbf{Input:} $P$, $R$\\
\hspace*{\algorithmicindent} \textbf{Output:} Reduced set of test paths $P_{reduced}$
\begin{algorithmic}[1]
\State $R_{uncovered} \gets R$
\State $P_{sorted} \gets$ $P$ sorted in descending way by number of covered requirements  
\State $P_{reduced} \gets$ an empty set of paths
\State $P_{sorted}$

\While {$R_{uncovered} \neq \emptyset \land P_{sorted} \neq \emptyset$}
    \State $p_{next} \gets$ first path from $P_{sorted}$
    \State $P_{sorted} \gets P_{sorted} \setminus \{p_{next}\}$
    \State $R_{covered} \gets$ requirements covered by $p_{next}$
     \If{$R_{covered} \neq \emptyset$}
        \State $R_{uncovered} \gets R_{uncovered} \setminus R_{covered}$
        \State $P_{reduced} \gets P_{reduced} \cup \{p_{next}\}$
    \EndIf
\EndWhile

\State \textbf{return} $P_{reduced}$
\Statex{\bf end function}
\end{algorithmic}
\end{algorithm}


\textbf{Chvatal} $P$ reduction variant that bases on Chvatal's algortihm for SCP \cite{Chvatal1979, Young2008} is defined by Algorithm \ref{alg:ChvatalsImprovedTestPathsReduction}. For performance reasons, the algorithm first filters $r \in R$ that are sub-path of other requirements $r \in R$ and removes paths $p \in P$ that do not cover any $r$. Chvatal's SCP algorithm uses $R$ as a universe and $P$ as a set of subsets where each path $p \in P$ represents a set of requirements $R_c \subseteq R$ that covers. During this SCP process, the reduced $P$ is generated.

\begin{algorithm}
\caption{Chvatal-SCP-based $P$ reduction}\label{alg:ChvatalsImprovedTestPathsReduction}
\hspace*{\algorithmicindent} \textbf{\bf Function: Chvatal}\\
\hspace*{\algorithmicindent} \textbf{Input:} $P$, $R$\\
\hspace*{\algorithmicindent} \textbf{Output:} Reduced set of test paths $P_{reduced}$
\begin{algorithmic}[1]

\State $R_{filtered} \gets$ $R \setminus R_{remove}$, where $R_{remove} = \{ r_1$ $|$ $r_1 \in R$, $r_1$ is not a sub-path of other $r_2 \in R$ and $r_1 \neq r_2$  $ \}$  

\State $P_{filtered} \gets$ $P$ from which all test paths that do not contain any $r \in R_{filtered}$ are removed

\State $P_{reduced} \gets $ reduce $P_{filtered}$ using Chvatal's SCP algorithm as defined in \cite{Chvatal1979} in such a way that $R$ is as an universe and $P$ is a set of subsets, where each $p \in P$ represents a $R_c \subseteq R$ that covers.

\State \textbf{return} $P_{reduced}$
\Statex{\bf end function}
\end{algorithmic}
\end{algorithm}

The next $P$ reduction variant, \textbf{Genetic Algorithm} \cite{whitley2012genetic} is employed to solve the SCP problem (Algorithm \ref{alg:GeneticTestPathReduction}). In the selection process, one population individual represents a $P_i \subseteq P$. Besides $P$ and $R$ on input, the algorithm employs a set of input configurations (further GA configuration), whose particular values were determined during the experiments as giving the best results. These are: \textit{initialProbabilityToSetGene=0.2}, \textit{probabilityToMutateOneGene=0.4}, \textit{probabilityToMutateZeroGene=0.6}, \textit{maxGenerations=100}, \textit{populationSize=30}, \textit{maxGenerationsWithoutImprovement=40}.

In this variant, we designed our fitness function, which is $fitness(individual) = maxCost - (uncovered(individual) \cdot ( |P| + 1 ) + |P_{individual}|)$, where $individual$ represents a $P_{individual} \subseteq P$,  $maxCost = |R| \cdot ( |P| + 1 ) + |P|$, and $uncovered(individual)$ is a count of $r \in R$ which are not a sub-path of any $p \in P_{individual}$. Crossover operation employed in the Genetic Algorithm $P$ reduction is further specified in Algorithm \ref{alg:Crossover}.



\begin{algorithm}
\caption{Genetic Algorithm based $P$ reduction}\label{alg:GeneticTestPathReduction}
\hspace*{\algorithmicindent} \textbf{\bf Function: Genetic Algorithm}\\
\hspace*{\algorithmicindent} \textbf{Input:} $P$, $R$, $populationSize$,\\ \hspace*{\algorithmicindent} $ initialProbabilityToSetGene,$\\ \hspace*{\algorithmicindent} $probabilityToMutateZeroGene,$\\
\hspace*{\algorithmicindent} $probabilityToMutateOneGene, maxGenerations,$\\
\hspace*{\algorithmicindent} $maxGenerationsWithoutImprovement$\\
\hspace*{\algorithmicindent} \textbf{Output:} Reduced set of test paths $P_{reduced}$
\begin{algorithmic}[1]

\State $R_{filtered} \gets$ $R$ from which filter requirements that are not a sub-path of any $p \in P$ are removed

\State initialize $population$
\Comment{Each individual is represented by an array of bits where each bit is set with probability $initialProbabilityToSetGene$}

\Do
    \State $offsprings \gets$ \textbf{Crossover(}$population$, $populationSize$, $|P|$, $probabilityToMutateZeroGene$, $probabilityToMutateOneGene$\textbf{)}
    \State $population \gets population \cup offsprings$
    \State $survivals_1 \gets$ ($populationSize / 10$) individuals from $population$ having the highest value of the fitness function
    \State $population \gets population \setminus survivals_1$
    \State $sutvivals_2 \gets$ ($populationSize - |survivals_1|$) survivals using a roulette wheel \cite{verma2014genetic} with same GA configuration from $population$
    \State $population \gets survivals_1 \cup survivals_2$
\doWhile {terminate criteria are not met} \Comment{terminate criteria are met when maximal number of generation $maxGenerations$ is achieved, best possible individual is found or there is no improvement for $maxGenerationsWithoutImprovement$ generations}

\State $P_{reduced} \gets$ test paths represented by selected best individuals

\State \textbf{return} $P_{reduced}$

\Statex{\bf end function}
\end{algorithmic}
\end{algorithm}

\begin{algorithm}
\caption{Crossover operation of Genetic Algorithm based $P$ reduction}\label{alg:Crossover}
\hspace*{\algorithmicindent} \textbf{\bf Function: Crossover}\\
\hspace*{\algorithmicindent} \textbf{Input:} $population, populationSize, chromosomeLegth$, \hspace*{\algorithmicindent} $probabilityToMutateZeroGene$,\\
\hspace*{\algorithmicindent} $probabilityToMutateOneGene$\\
\hspace*{\algorithmicindent} \textbf{Output:} $offsprings$

\begin{algorithmic}[1]

\State $offsprings \gets$ init empty set of individuals

\For{$n=1~to~\lfloor populationSize/3 \rfloor$}
    \State Choose $parent_1$ and $parent_2$ using the roulette wheel using the GA configuration
    \State $point_1, point_2 \gets$ generate two random points such that $0 < point_1 < point_2 < chromosomeLength$
    \State Create $offspring_1$ and $offspring_2$ by combination of parents' chromosomes, such that $offspring_1$ has zero to $point_1^{~th}$ gene of $parent_1$, $point_1^{~th}$ to $point_2^{~th}$ gene of $parent_2$ and $point_2^{~th}$ to the last gene of $parent_1$. For $offspring_2$ do the same process, only switch $parent_1$ and $parent_2$.
    \State Mutate one random gene of $offspring_1$ and $offspring_2$ with probability $probabilityToMutateOneGene$ or $probabilityToMutateZeroGene$ if the gene is set or not respectively 
    
    \State $offsprings \gets offsprings \cup \{offspring_1, offspring_2\}$
\EndFor

\State \textbf{return} $offsprings$

\Statex{\bf end function}
\end{algorithmic}
\end{algorithm}


The last variant to compare is $P$ reduction done via \textbf{Simulated Annealing} as specified by Algorithm \ref{alg:SimulatedAnnealingTestPathReduction}. In the algorithm, we have used a geometric annealing schedule. Hence, besides $P$ and $R$, the algorithm accepts the $\alpha$ coefficient for this schedule and, based on the experiment results, $\alpha=0.8$ was used.

\begin{algorithm}
\caption{Simulated Annealing $P$ reduction}\label{alg:SimulatedAnnealingTestPathReduction}
\hspace*{\algorithmicindent} \textbf{\bf Function: Simulated Annealing}\\
\hspace*{\algorithmicindent} \textbf{Input:} $P$, $R$, coefficient $\alpha$ for geometric annealing schedule\\
\hspace*{\algorithmicindent} \textbf{Output:} $P_{reduced}$

\begin{algorithmic}[1]

\State $point_{actual} \gets$ a random point in the solution space \Comment{Solution is represented by an array of bits of size $|P|$. A point defines $p \in P$, that are used in the solution, if $n^{th}$ bit is set then $n^{th}$ path of $P$ is used. $paths(point)$ denotes a $P_{point} \subseteq P$ that is represented by $point$. Further, $covered(paths)$ denotes a $R_{covered} \subseteq R$ such that each $r \in R_{covered}$ is a sub-path of a $p \in P$}

\State $point_{best} \gets point_{actual}$

\State $t \gets |P|^{2.2} \cdot 1.5$

\Do
\State $t' \gets t$
    \Do
        \State $point_{neighbour} \gets$ $point$ with a random bit switched
        \State $delta \gets$ $energy(point_{neighbour}) - energy(point)$ 
        \Comment{$energy(point) = |R \setminus covered(paths(point))| \cdot (|P| + 1) + |paths(point)|$}
        
        \If{$delta < 0$}
            \State $point_{actual} \gets point_{neighbour}$
            \If{$energy(point_{neighbour}) < energy(point_{best})$}
                \Comment{Check whether the new solution is the new best one}
                \State $point_{actual} \gets point_{neighbour}$
            \EndIf
        \ElsIf{$Random(0,1) < Exp(-delta/t)$}
            \Comment{$Exp(x)$ returns the natural exponential of x and $Random(a,b)$ returns random number $n \in [a,b]$ of uniform distribution}
            \State $point_{actual} \gets point_{neighbour}$ 
        \EndIf
    \doWhile{the loop is iterated $\ceil{t}$-times where $t$ is actual temperature} \Comment{Equilibrium is reached}
    
    \State $t \gets \alpha \cdot t$ \Comment{Geometric cooling schedule is used}
    
\doWhile{$t<0.000001$} \Comment{Is frozen}

\State $P_{reduced} \gets paths(point_{best})$

\State \textbf{return} $P_{reduced}$

\Statex{\bf end function}
\end{algorithmic}
\end{algorithm}

The \textbf{N-Switch Reduction strategy} bases on established \textit{N-Switch Coverage} concept used in FSM testing\cite{vroon2013tmap,pol2002software,aalst2008tmap} and is defined by Algorithm \ref{alg:NSwitchReduction}. The algorithm accepts the same inputs as the PSMT strategy (see Algorithm \ref{alg:GenerateTestPaths}) and produces $P$ as an output. The algorithm first generates all paths in $\boldsymbol{G}$ of length $n$, where $minLength \leq n \leq maxLength$, which are put to the initial $P$. Then, the algorithm reduces $P$ to keep only paths that starts in $\{v_s\} \cup V_{ts}$ and ends in some vertex from $V_e \cup V_{te}$.  Then, paths of $p$ are further analyzed if more paths start in a particular vertex from $\{v_s\} \cup V_{ts}$ and if so, only one of these paths is kept in $P$.

\begin{algorithm}
\caption{Generate test paths for SUT model by N-Switch Reduction strategy}
\label{alg:NSwitchReduction}
\hspace*{\algorithmicindent} \textbf{Function: N-Switch Reduction}\\
\hspace*{\algorithmicindent} \textbf{Input:} $\boldsymbol{G}$, $minLength$, $maxLength$, $minPriority$, $maxPriority$, $testCoverage$\\
\hspace*{\algorithmicindent} \textbf{Output:} Set of test paths $P$
\begin{algorithmic}[1]
\State $P_{full} \gets \emptyset$; $P \gets \emptyset$ \Comment{empty set of paths}
\For{$e \in E$}
    \State $P_e \gets$ all paths in $\boldsymbol{G}$ such that for each $p \in P_e$ $minLength \leq length(p) \leq minLength$ and $e$ is the first edge in $p$
    \State $P_{full} \gets P_{full} \cup P_e$
\EndFor
\State $R \gets$ generate requirements for $testCoverage$ for vertices $v \in V$ with $minPriority \leq priority(v) \leq maxPriority$ and edges $e \in E$ with $minPriority \leq priority(e) \leq maxPriority$ by specification in Section \ref{sec:test_coverage_criteria}.
\For{each $p \in P_{full}$}
    \State $R_p \gets$ requirements from $R$ which are sub-path of $p$
    \If{$R_p \neq \emptyset$}
        \State $R \gets R \setminus R_p$; $P \gets P \cup \{p\}$
        \If{$R = \emptyset$}
            \State \textbf{break the loop}
        \EndIf
    \EndIf
\EndFor
\State \textbf{return} $P$
\Statex{\bf end function}
\end{algorithmic}
\end{algorithm}


\section{Experiments}

\begin{figure*}
    \centering
    \includegraphics[width=18cm]{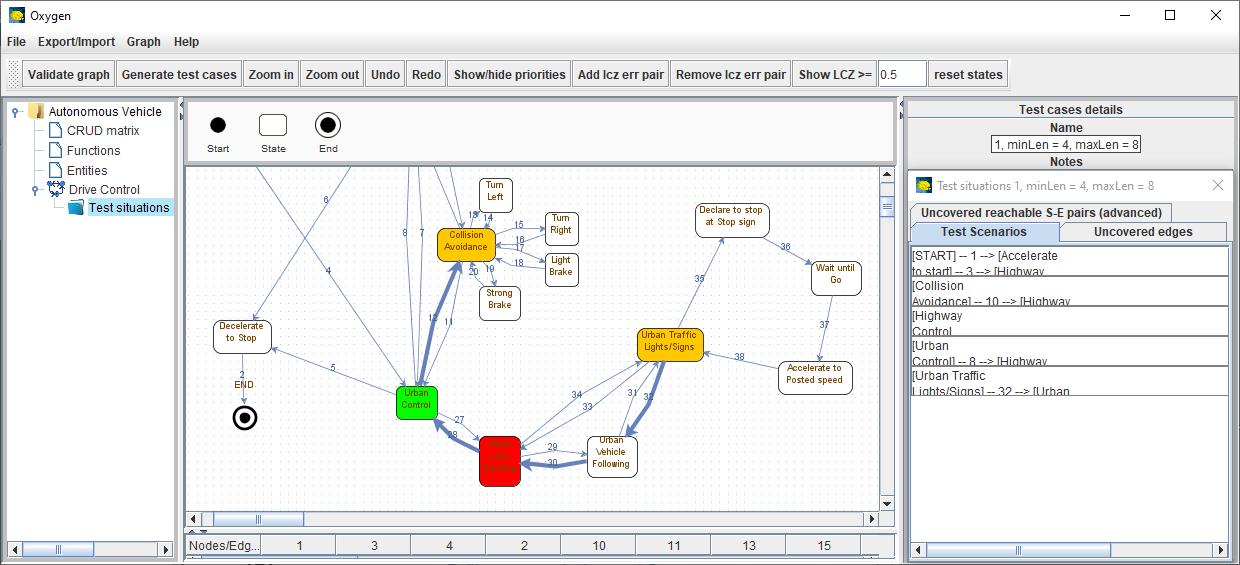}
    \caption{Example of Oxygen editor with visualization of a test path in a SUT model.}
    \label{fig:oxygen_screenshot}
\end{figure*}

All presented algorithms are implemented in the developer version of the \textit{Oxygen}\footnote{http://still.felk.cvut.cz/oxygen/} platform \cite{bures2015pctgen,bures2017prioritized}.
The platform allows for the creation of the SUT model in its graphical editor. It also allows to import and export of the SUT models in open XML-based formats. The selected algorithms can generate test path sets for a set of SUT models. Test paths can then visualized in the model and exported. An example of Oxygen user interface with visualized test path in a SUT model is given in Figure \ref{fig:oxygen_screenshot}. 
SUT model states from $V_{ts}$, $|V_{te}|$ and $|V_{ts} \cap V_{te}|$ are highlighted by green, red and orange color, respectively. The developer version of the platform allows for batch execution of the present algorithms on multiple SUT models and export of summary results for further analysis and evaluation.

\subsection{Experiment set-up and method}

For the experiments, we used 180 problem instances $\boldsymbol{G}$. These instances are composed of: (1) state machine models of real industrial project, (2) modifications of the industrial real models, and (3) problem instances generated artificially. Regarding the industrial FSMs, six models describing various parts of tested Skoda cars were used. The Skoda Auto testing team created these models in a special Skoda version of the Oxygen tool supporting $\boldsymbol{G}$. These models were further modified by adding cycles to an FSM, adding possible test starts and test ends, adding or removing a state, and adding and removing a transition. In this paper, by a cycle, we mean a configuration of edges that allow non-empty trails in which only the first and last vertices are equal. The result was 24 problem instances. Another 12 problem instances were created by the same method for FSMs for data transmission protocols and mission management in the \textit{Digital Triage Assistant} project\footnote{https://www.natomultimedia.tv/app/asset/656263}, a joint project of CTU in Prague, Johns Hopkins University, University of Defence, NATO ACT IH and other partners. Non-disclosure agreements and confidentiality restrictions of both projects allow making publicly available only abstracted topology of FSMs without the names of states and transitions.


To create enough variety of possible FSMs, we generated an additional set of 144 problem instances by a specialized tool, developed as a master thesis by Richard Sadlon\footnote{https://dspace.cvut.cz/handle/10467/97079}, CTU in Prague. The tool generates $\boldsymbol{G}$ problem instances by expected properties of the graph entered as an input. In this process, $|V|$, $|E|$, number of $\boldsymbol{G}$ cycles, $|V_{ts}|$,  $|V_{te}|$, $|V_{ts} \cap V_{te}|$, and $|V_{e}|$ can be specified.


Table \ref{tab:problem_instances} presents the selected properties of all problem instances used in the experiments. In Table \ref{tab:problem_instances}, $cycles$ denotes number of $\boldsymbol{G}$ cycles, $avg \: cycle \: length$ denotes average length of these cycles, $parallel \: edge \: groups$ denotes number of groups of parallel edges present in $\boldsymbol{G}$ (in these groups, edges start and end in the same vertex), $parallel \: edges$ denotes total number of parallel edges in $\boldsymbol{G}$, $avg \: D+$ denotes average node incoming degree, $avg \: D-$ denotes average node outgoing degree, and $avg \: D$ denotes average node degree. Further, $|V_{ts} \cap V_{te}|$ denotes the state in which a test path can both start and end.

Regarding the priorities defined in the SUT model, $priority(v)$ as well as $priority(e)$ were set to range 0 to 3 for all $v \in V$, resp. $e \in E$, for all problem instances. Then, $minPriority$ was set to 2 and $maxPriority$ was set to 3 in a unified way for the experiment. In Table \ref{tab:problem_instances}, $priority \: vertices$ denotes number of $v \in V$ for which $minPriority \leq priority(e) \leq maxPriority$ and \ref{tab:problem_instances}, $priority \: edges$ denotes number of $e \in V$ for which $minPriority \leq priority(v) \leq maxPriority$.

To evaluate the effectiveness of the generated $P$ to detect defects, the problem instances were extended by fictional defects of two types. \textit{Type 1} Defect is present at an edge $e \in E$ representing FSM transition and is considered to be activated when a $p \in P$ visits $e$. In Table \ref{tab:problem_instances}, number of these defects in problem instances is denoted as $type \: 1 \: defects$. 

\textit{Type 2} defect is present at two edges $e_1 \in E$ and $e_2 \in E$, where there is a path in $\boldsymbol{G}$ from $e_1$ to $e_2$. To consider \textit{Type 2} defect to be activated, a $p \in P$ visits $e_1$ and then visits $e_2$. \textit{Type 2} Defects simulate data consistency defects, when a transition $e_1$ induces an internal inconsistency to a SUT and other transition ($e_2$) causes a defective behavior of the SUT.

In Table \ref{tab:problem_instances}, the number of \textit{Type 2} defects in problem instances is denoted as $type \: 2 \: defects$ and average distance between $e_1$ and $e_2$ in number of edges is denoted as $e_1 \: to \: e_2 \: avg \: distance$.

\begin{table}[h!]
\caption{Properties of problem instances used in the experiments.}
\centering
\begin{tabular}{|c|c|c|c|c|}
\hline
 \textbf{$\boldsymbol{G}$ property} & \textbf{Mean} & \textbf{Median} & \textbf{MIN} & \textbf{MAX}\\
\hline

$|V|$ &	20.33	&	15	&	10	&	57
 \\ \hline
$|E|$ &	34.81	&	35	&	19	&	95
 \\ \hline
$cycles$ &	3.23	&	3	&	0	&	18
 \\ \hline
$avg \: cycle \: length$ &	9.27	&	8	&	0	&	31
 \\ \hline
$|V_{e}|$ &	2.09	&	1	&	1	&	21
 \\ \hline
$parallel \: edges$ &	0.53	&	0	&	0	&	18
 \\ \hline
$parallel \: edge \: groups$ &	0.27	&	0	&	0	&	9
 \\ \hline
$avg \: D+$ &	1.77	&	1.53	&	1.04	&	2.33
 \\ \hline
$avg \: D-$ &	1.77	&	1.53	&	1.04	&	2.33
 \\ \hline
$avg \: D$ 	&	3.55	&	3.07	&	2.08	&	4.67
 \\ \hline
$|V_{ts}|$ &	2.33	&	2	&	1	&	17
 \\ \hline
$|V_{te}|$ &	2.59	&	2	&	1	&	25
 \\ \hline
$|V_{ts} \cap V_{te}|$ &	1.14	&	1	&	0	&	6
 \\ \hline
$priority \: nodes$ &	5.66	&	5	&	0	&	21
 \\ \hline
$priority \: edges$ &	9.75	&	9	&	2	&	34
 \\ \hline
$type \: 1 \: defects$ &	7.00	&	6	&	1	&	28
 \\ \hline
$type \: 2 \: defects$ &	6.26	&	6	&	1	&	27
 \\ \hline
$e_1 \: to \: e_2 \: avg \: distance$ &	2.69	&	2.50	&	0.67	&	6.60
 \\ \hline

\end{tabular}
\label{tab:problem_instances}
\end{table}

We evaluated the following properties of $P$: $|P|$, total length of all $p \in P$ in number of edges (denoted as $steps$), average length of all $p \in P$ in number of edges (denoted as $avg\_steps$), number of unique edges in all $p \in P$ (denoted as $unique\_steps$). We also analyzed $ut = \frac{len}{unique}$ that expresses how many non-unique FSM transitions ($\boldsymbol{G}$ edges) need to repeat in a test path to test the unique transitions. Higher $ut$ indicates higher "edge duplication" in $P$.

To evaluate the defect detection potential of $P$, we analyzed the number of simulated defects of \textit{Type 1} and \textit{Type 2} that were activated by a $p \in P$, denoted as $Type \: 1 \: activated$ and $Type \: 2 \: activated$, respectively. The last property to evaluate is averaged number of simulated defects activated by one test path step, defined as $eff_1 = \frac{Type \: 1 \: activated}{steps}$ and $eff_2 = \frac{Type \: 2 \: activated}{steps}$ for \textit{Type 1} and \textit{Type 2} simulated defects, respectively.

We ran all presented algorithms for both \textit{PSMT Basic Coverage} and \textit{PSMT Extended Coverage} criteria. In each of these criteria, we present and discuss the results for two test path length ranges: $minLength=2, maxLength=6$ and $minLength=4, maxLength=8$. If no $P$ was returned for a particular problem instance for the individual configuration of test path length range and the coverage criterion by any of the compared algorithms, the result record was not taken into account for all algorithms for this problem instance. Because of the non-deterministic nature of the Genetic Algorithm and Simulated Annealing test path reduction variants of the PSMT strategy, we ran all computations three times and averaged the results.

\subsection{Results}

Table \ref{tab:properties_of_P_2_to_6} presents the averaged results for all problem instances for all compared algorithms and both coverage criteria for the test path length range $minLength=2, maxLength=6$. The major properties we further analyze and discuss are $steps$, $eff_1$ and $eff_2$. The best results for these properties are marked by bold and the second-best result by italics in the data. Regarding $Type \: 1 \: activated$, $Type \: 2 \: activated$, $eff_1$ and $eff_2$ in Table \ref{tab:properties_of_P_2_to_6}, particular simulated defect can be activated more times by test paths from $P$ and these multiple activation are included in the results.  

For \textit{PSMT Basic Coverage}, the Sort variant of PSMT produced $P$ with the lowest number of total steps (13.87 on average), followed by Simulated Annealing variant (14.28 on average). This is more than a twofold difference compared to the No reduction variant and approximately one-third difference to N-Switch Reduction serving as a baseline. The most important properties to evaluate are $eff_1$ and $eff_2$. The best average value of $eff_1$ was achieved in $P$ generated by the Sort variant of PSMT (0.313 defects per one test path step), followed by Simulated Annealing variant (0.304). This is 20\% better result than the baseline N-Switch Reduction. In the case of $eff_2$, N-Switch Reduction (average $eff_2=0.04$) outperforms Sort and Chvatal PSMT variants, the second-best performers, by approximately 25\%. However, a relatively small portion of $Type_2$ defects is detected by generated $P$ for all algorithms, approx. ten-times less than for \textit{Type 1} defects. 

Analyzing the same test coverage criteria for $minLength=4, maxLength=8$ test path length range (see Table \ref{tab:properties_of_P_4_to_8}), Genetic Algorithm variant of PSMT generated $P$ with the best $steps$ (18.55), followed by Sort variant (18.58). Baseline N-Switch Reduction produced $P$ with the number of steps 56\% higher than the Genetic Algorithm variant. Regarding $eff_1$, the Genetic Algorithm and Sort variants yield the best result (0.297), 30\% better than N-Switch Reduction. For \textit{Type 2} defects ($eff_2$), N-Switch Reduction outperforms the other algorithms, and the difference to the second-best performer, the Simulated Annealing PSMT variant, is 10\%.

\begin{table*}[h!]
\caption{Properties of generated $P$ for $minLength=2, maxLength=6$, average value for all problem instances.}
\centering
\begin{tabular}{|c|c|c|c|c|c|c|c|}
\hline

& \multicolumn{6}{|c|}{\textbf{{PSMT strategy variant}}} 
& \textbf{N-Switch Reduction}\\
\cline{2-7}

 \textbf{$P$ property} & \textbf{No reduction} & \textbf{Random} & \textbf{Sorted} & \textbf{Chvatal} & \textbf{Genetic Algorithm} & \textbf{Simulated Annealing} & \textbf{\textit{(baseline)}}\\
\hline

\multicolumn{8}{|l|}{\textbf{\textit{{PSMT Basic Coverage}}}} \\
\hline
$steps$  &	29.18	&	15.75	&	\textbf{13.87}	&	14.99	&	15.63	&	\textit{14.28}	&	22.05
 \\ \hline
$|P|$   &	9.30	&	4.83	&	4.15	&	4.54	&	4.81	&	4.31	&	4.68
 \\ \hline
$avg\_steps$   &	3.28	&	3.43	&	3.54	&	3.51	&	3.55	&	3.54	&	4.66
	\\ \hline
$unique\_steps$   &	13.97	&	11.89	&	11.31	&	11.57	&	11.64	&	11.43	&	14.26
	\\ \hline
$ut$   &	2.12	&	1.27	&	1.18	&	1.23	&	1.20	&	1.18	&	1.45
 \\ \hline
$Type \: 1 \: activated$   &	5.35	&	4.55	&	4.36	&	4.45	&	4.47	&	4.39	&	5.44
 \\ \hline
$Type \: 2 \: activated$   &	0.58	&	0.44	&	0.45	&	0.48	&	0.46	&	0.44	&	1.00
 \\ \hline
$eff_1$   &	0.182	&	0.282	&	\textbf{0.313}	&	0.297	&	0.298	&	\textit{0.304}	&	0.259
 \\ \hline
$eff_2$  &	0.020	&	0.028	&	\textit{0.032}	&	\textit{0.032}	&	0.030	&	0.030	&\textbf{0.040}
 \\ \hline

\multicolumn{8}{|l|}{\textbf{\textit{{PSMT Extended Coverage}}}} \\
\hline
$steps$   &	141.77	&	58.04	&	\textbf{51.29}	&	\textit{56.34}	&	95.82	&	78.01	&	66.31
 \\ \hline
$|P|$   &	40.09	&	15.13	&	12.84	&	14.68	&	27.15	&	21.26	&	13.36
 \\ \hline
$avg\_steps$   &	3.61	&	3.89	&	4.07	&	3.87	&	3.95	&	3.97	&	4.61
	\\ \hline
$unique\_steps$   &	18.76	&	18.68	&	18.62	&	18.66	&	18.67	&	18.66	&	19.11
	\\ \hline
$ut$   &	6.73	&	2.33	&	2.05	&	2.26	&	2.98	&	2.53	&	2.52
 \\ \hline
$Type \: 1 \: activated$   &	7.22	&	7.20	&	7.18	&	7.20	&	7.20	&	7.20	&	7.36
 \\ \hline
$Type \: 2 \: activated$   &	1.36	&	1.36	&	1.34	&	1.36	&	1.36	&	1.36	&	1.94
 \\ \hline
$eff_1$  &	0.064	&	0.192	&	\textbf{0.215}	&	0.194	&	0.196	&	\textit{0.203}	&	0.185
 \\ \hline
$eff_2$  &	0.009	&	0.026	&	\textit{0.030}	&	0.027	&	0.026	&	0.027	&	\textbf{0.031}
 \\ \hline

\end{tabular}
\label{tab:properties_of_P_2_to_6}
\end{table*}

For \textit{PSMT Extended Coverage} criterion and $minLength=2, maxLength=6$ length range (second part of Table \ref{tab:properties_of_P_2_to_6}), the best average $steps$ is achieved by Sort PSMT variant (51.29), followed by Chvatal variant (56.3). The result of the N-Switch Reduction baseline is 30\% worse compared to the Sort variant. In terms of $eff_1$, the Sort variant yields the best result (0.215), followed by the Simulated Annealing variant (0.203). The result of the Sort variant is better by 16\% than for the N-Switch Reduction baseline. Regarding $eff_2$, no significant difference between N-Switch Reduction and the second-best algorithm, the Sort variant of PSMT is present in the data.

For the $minLength=4, maxLength=8$ length range, the Sort PSMT variant gives the best average value of $steps$ (61.03), and the baseline N-Switch Reduction result is 40\% higher (85.85). The second-best performer in $steps$ is the Chvatal variant (63.48). For $eff_1$, Sort PSMT variant performed best with 0.198 defects per one $P$ steps, a 20\% difference to N-Switch Reduction baseline (0.164). The second-best performers were Chvatal and Simulated Annealing variants, $eff_1=0.186$. No significant differences were observed for $eff_2$ among the compared algorithms. 

Generally, the differences in $eff_1$ between the best algorithm and N-Switch Reduction baseline are higher for \textit{PSMT Basic Coverage} than for \textit{PSMT Extended Coverage}. For $eff_2$, N-Switch Reduction outperformed other algorithms for \textit{PSMT Basic Coverage}. No significant differences were observed for \textit{PSMT Extended Coverage} criterion.


\begin{table*}[h!]
\caption{Properties of generated $P$ for $minLength=4, maxLength=8$, average value for all problem instances.}
\centering
\begin{tabular}{|c|c|c|c|c|c|c|c|}
\hline

& \multicolumn{6}{|c|}{\textbf{{PSMT strategy variant}}} 
& \textbf{N-Switch Reduction}\\
\cline{2-7}

 \textbf{$P$ property} & \textbf{No reduction} & \textbf{Random} & \textbf{Sorted} & \textbf{Chvatal} & \textbf{Genetic Algorithm} & \textbf{Simulated Annealing} & \textbf{\textit{(baseline)}}\\
\hline

\multicolumn{8}{|l|}{\textbf{\textit{{PSMT Basic Coverage}}}} \\
\hline
$steps$  &	45.92	&	20.86	&	\textit{18.58}	&	20.33	&	\textbf{18.55}	&	20.07	&	29.02
 \\ \hline
$|P|$ &	9.73	&	4.37	&	3.85	&	4.24	&	3.85	&	4.21	&	4.40
 \\ \hline
$avg\_steps$  &	4.89	&	5.05	&	5.14	&	5.08	&	5.14	&	5.13	&	6.45
	\\ \hline
$unique\_steps$   &	16.58	&	14.26	&	13.58	&	14.04	&	13.63	&	13.79	&	16.03
	\\ \hline
$ut$   &	2.81	&	1.41	&	1.32	&	1.39	&	1.30	&	1.33	&	1.69
 \\ \hline
$Type \: 1 \: activated$   &	6.36	&	5.46	&	5.26	&	5.43	&	5.30	&	5.35	&	6.14
 \\ \hline
$Type \: 2 \: activated$   &	1.11	&	0.89	&	0.82	&	0.84	&	0.81	&	0.84	&	1.33
 \\ \hline
$eff_1$ &	0.144	&	0.271	&	\textbf{0.297}	&	0.280	&	\textbf{0.297}	&	\textit{0.295}	&	0.229
 \\ \hline
$eff_2$ &	0.025	&	0.042	&	0.042	&	0.042	&	0.042	&	\textit{0.043}	&	\textbf{0.047}
 \\ \hline

\multicolumn{8}{|l|}{\textbf{\textit{{PSMT Extended Coverage}}}} \\
\hline
$steps$  &	203.10	&	66.38	&	\textbf{61.03}	&	\textit{63.48}	&	115.35	&	95.50	&	85.85
 \\ \hline
$|P|$  &	42.33	&	13.53	&	12.32	&	12.89	&	24.76	&	20.16	&	12.41
 \\ \hline
$avg\_steps$   &	5.02	&	5.19	&	5.33	&	5.25	&	5.25	&	5.25	&	6.50
	\\ \hline
$unique\_steps$  &	20.22	&	20.06	&	19.99	&	20.07	&	20.06	&	20.05	&	20.40
	\\ \hline
$ut$ &	9.25	&	2.51	&	2.33	&	2.42	&	3.43	&	2.93	&	3.04
 \\ \hline
$Type \: 1 \: activated$   &	7.76	&	7.71	&	7.69	&	7.73	&	7.72	&	7.71	&	7.83
 \\ \hline
$Type \: 2 \: activated$  &	1.96	&	1.85	&	1.85	&	1.85	&	1.89	&	1.88	&	2.55
 \\ \hline
$eff_1$ &	0.045	&	0.183	&	\textbf{0.198}	&	\textit{0.186}	&	0.184	&	\textit{0.186}	&	0.164
 \\ \hline
$eff_2$ &	0.009	&	0.034	&	\textbf{0.037}	&	\textit{0.035}	&	0.033	&	0.034	&	\textit{0.035}
 \\ \hline

\end{tabular}
\label{tab:properties_of_P_4_to_8}
\end{table*}

\section{Discussion}

Various properties of $P$ can be evaluated. Property $steps$ can be considered as a reasonable proxy for the effort to execute or automate the tests. When the goal is to satisfy the given test coverage criteria, the reduction of $steps$ makes good sense from a test effort viewpoint. However, the reduction of $steps$ has to be analyzed in the context of defects that can be found in SUT by $P$. More extensive $P$ increases the chances of activating defects present in SUT, and reducing $steps$ can decrease this chance. Hence, $eff_1$ and $eff_2$ values based on fictional defects in FSM and their activation by $P$ have to be analyzed as more objective indicators.

In the presented data, one distribution of \textit{Type 1} and \textit{Type 2} was used for each of the 180 problem instances. With the other distribution of fictional defects, the results might differ. Moreover, the preferred test path length ($<MinLength-MaxLength>$ interval) influences the results, as can be seen in the comparison of Tables \ref{tab:properties_of_P_2_to_6} and \ref{tab:properties_of_P_4_to_8}.

Despite the general good performance of Sort variant of PSMT in $steps$ and $eff_1$, also Genetic Algorithm, Simulated Annealing and Chvatal set cover variants performed well in individual combinations of parameters, test coverage criteria and test path length range (see Tables \ref{tab:properties_of_P_2_to_6} and \ref{tab:properties_of_P_4_to_8}). Moreover, the baseline N-Switch Reduction gave the best results for $eff_2$ in three out of four evaluated combinations of test coverage criteria and test path length range. However, in interpreting this result, we need to consider that only a relatively small portion of $Type_2$ defects is detected by generated $P$ for all algorithms. Here, the average number of defects detected by one $P$ step is approx. tenfold lower than this number for \textit{Type 1} defects. Hence, the classical FSM-based technique, as discussed in this paper, shows potentially ineffective to detect $Type_2$ defects and shall be replaced by some specialized data consistency testing techniques, e.g., \cite{bures2020dynamic}.   


To conclude, no "universal clear winner" can be identified among the analyzed algorithms. Sort variant of PSMT gives the best results for $steps$ and detection of \textit{Type 1} defects on average. However, it has not produced the $P$ with the best $eff_1$ for all combinations of the problem instance, test coverage criteria, and test path length range in the experiments. N-Switch Reduction was the best option for \textit{Type 2} defects for \textit{PSMT Basic Coverage} criterion. But also, this result was not achieved for all combinations of the problem instance, test coverage criteria, and test path length range. For \textit{PSMT Extended Coverage} criterion, no significant difference was observed for the presented experiment configuration.

Hence, a combination of the presented algorithms with a portfolio strategy (e.g., \cite{bures2019employment}) seems like the best option to yield close-to-optimum $P$. Such a strategy would compute $P_1..P_n$ for particular $\boldsymbol{G}$ by individual algorithms and selection of $P$ having the best property which is given in input.

\section{Related work}

The number of algorithms generating a set of test paths from an FSM-based model was published in the literature \cite{lee1996principles,dias2007survey,aggarwal2012test,ammann2016introduction, kalaji2009generating,mouchawrab2010assessing}. However, much fewer works exist on the problem as defined in this paper. Hence, we analyze algorithms and approaches containing elements that overlap our problem and can potentially be utilized for its solution.

Liu and Xu proposed a comparable approach to generate a test path set for FSM  \cite{MTToolAToolForSoftwareModelingAndTestGeneration}. As a SUT model, RFSM is employed. RFSM extends FSM with a special notation of labels that give the RFSM the ability to model more details \cite{ANewApproachToGeneratingHighQualityTestCases}. An algorithm is used to transform RFSM into a regular expression and generate the test paths. In another proposal by Alava \textit{et.~al.}, FSM-based test path generation is used to generate automated tests for web applications \cite{AutomaticValidationOfJavaPageFlowsUsingModelBasedCoverageCriteria}. The proposal practically supports Node Coverage Edge Coverage criteria in FSM.

Devroey~\textit{et~al.} generate test suites for software product lines using a problem model that bases on FSM \cite{AbstractTestCaseGenerationForBehaviouralTestingOfSoftwareProductLines}. Authors use a branch and bound approach and use heuristics for efficient test path search. The algorithm explores the graph using a prioritization mechanism. Prioritization score is evaluated using an accessibility matrix computed by a modified Warshall algorithm. Klalil and Labiche presented an approach for generating test paths from FSM \cite{StateBasedTestsSuitesAutomaticGenerationTool} that supports the Round Trip Coverage criterion. Alternative test coverage criteria such as Random, Depth Traversal and Breadth Traversal are discussed as well and compared.

In the alternative approaches, Carvalho and Tsuchiya exploit model checking to generate test paths for SUT parts described as FSMs, as model checkers can generate counterexamples as a proof when a model does not satisfy the specification \cite{CoverageCriteriaForStateTransitionTestingAndModelCheckerBasedTestCaseGeneration}. The proposal practically supports Node Coverage, Edge Coverage and Edge-pair Coverage criteria. Kilincceker~\textit{et al.} proposed a test path generation method from an FSM, which is further conversed to a regular expression \cite{Kilincceker2019}. In this approach, a Context Table is used, and the source code of the algorithms can be obtained freely \cite{KilinccekerGithubMBIT4SW}. Kilincceker~\textit{et al.} also presented an approach for generating test paths using an FSM-based SUT model that is derived from a specification in Hardware Description Language (HDL) language \cite{RegularExpressionBasedTestSequenceGenerationForHDLProgramValidation} and further transformed to a regular expression, which is an extension of the model by Liu et al. mentioned in \cite{AStudyForExtendedRegularExpression-basedTesting}. To obtain the test paths, the regular expression is parsed into a syntax tree from which the test paths are finally generated using an algorithm specified in the study. Fazli and Mohsen explore the generation of prime test paths  \cite{Fazli2019ATimeAndSpaceEfficientCompositionalMethodForPrimeAndTestPathsGeneration} and compare three approaches for prime path generation for FSM. Their proposal outperforms the compared alternatives in terms of memory consumption and processing time.



Despite the fact that FSM testing is a well-established subarea of the system testing discipline, no work we have found is concurrently and directly addressing (1) the possibility to explicitly set the start and end of a test path in a SUT model, (2) the possibility to prioritize the states and transitions of FSM to be present in the test paths, and, (3) possibility to determine expected length range of generated test paths \cite{aggarwal2012test,kalaji2009generating,mouchawrab2010assessing,dias2007survey}.

\section{Conclusion}

The paper presents a PSMT strategy to generate test path sets from an FSM-based SUT model, which concurrently allows for (1) expression of FSM states in which test paths can start and end, (2) selection of FSM priority states and transitions that must be included in a set of test paths, whereas the other can be neglected, and (3) specification of a length range that the generate test paths must fit in. These requirements result from the analysis of the FSM-based testing process in four large companies, Rockwell, Siemens, Skoda Auto and Electrolux. A concurrent combination of requirements (1)-(3) gives the technique good flexibility in real industrial testing processes. Depending on the application of the FSM, the strategy is applicable for the testing of both functional and non-functional software requirements.

Since addressing these three requirements concurrently is under-explored in the literature, the presented findings are useful for further research in this direction. Six variants of the PSMT strategy were presented and compared with an N-Switch Reduction baseline based on the established N-switch coverage concept.

More extensive experiments and measurements have to be done in the future to draw more conclusions, as the results might vary by the distribution of artificial defects and the distribution of priority FSM states and transitions in FSMs employed as problem instances in the experiments. Out of the compared alternatives, Sort variant, Simulated Annealing, and Genetic Algorithm variants give promising results to be elaborated further. Out of these, the Sort variant was the best performer overall. However, the results indicate that a universal algorithm performing the best way for all possible problem instances would be difficult to formulate. Instead, combining a few best performers with a portfolio strategy is likely to obtain the best results.

\section*{Acknowledgment}

The project is supported by CTU in Prague internal grant SGS20/177/OHK3/3T/13 “Algorithms and solutions for automated generation of test scenarios for software and IoT systems.” The authors acknowledge the support of the OP VVV funded project CZ.02.1.01/0.0/0.0/16\_019 /0000765 “Research Center for Informatics.” Bestoun S. Ahmed has been supported by the Knowledge Foundation of Sweden (KKS) through the Synergi Project AIDA - A Holistic AI-driven Networking and Processing Framework for Industrial IoT (Rek:20200067).

\bibliographystyle{IEEEtran}
\bibliography{REFERENCES}

\end{document}